\documentclass[onecolumn]{aa}
\def\gsim{ \lower .75ex \hbox{$\sim$} \llap{\raise .27ex \hbox{$>$}} }
\def\lsim{ \lower .75ex\hbox{$\sim$} \llap{\raise .27ex \hbox{$<$}} }
\def\observed{{\it observed}\ }
\def\intrinsic{{\it intrinsic}\ }
\usepackage{graphicx}

\authorrunning{Guetta \& Piran}
\titlerunning{Short duration GRBs}

\begin{document}
\title{The BATSE-{\it Swift} luminosity and redshift distributions of short-duration GRBs}
\author{Dafne Guetta\inst{1}
\and Tsvi Piran\inst{2}}

\institute{ Osservatorio astronomico of Rome v. Frascati 33 00040
Monte Porzio Catone, Italy  \and  Racah Institute for Physics, The
Hebrew University, Jerusalem 91904, Israel}
\date{To be determined}

\abstract{ We compare the luminosity function and rate inferred from
the BATSE short hard bursts (SHBs) peak flux distribution with the
redshift and luminosity distributions of SHBs observed by {\it
Swift}/HETE II. While the {\it Swift}/HETE II SHB sample is
incompatible with SHB population that follows the star formation
rate, it is compatible with a SHB rate that reflects a distribution
of delay times after the SFR. This would be the case if SHBs are
associated with binary neutron star mergers.  The available data
allows, however,  different interpretations. For example,  a
population whose rate is independent of the redshift fits the data
very well. The implied SHB rates that we find range from $\sim 8$ to
$\sim 30h_{70}^3$Gpc$^{-3}$yr$^{-1}$. This rate, which is comparable
to the rate of neutron star mergers estimated from statistics of
binary pulsars, is a much higher rate than what was previously
estimated. We stress that our analysis, which is based on \observed
short hard burst is limited to bursts with luminosities above
$10^{49}$erg/sec. Weaker burst may exist but if so they are hardly
detected by BATSE of {\it Swift} and hence their rate is very weakly
constrained by current observations.

} \maketitle

\section{Introduction}

It has been known since the early nineties that  gamma-ray bursts
(GRBs) are divided to two subgroups of long and short according to
their duration: $T_{90}\lsim 2$sec (Kouveliotou et al. 1993).
Short bursts are also harder than long ones (Dezalay et al. 1996,
Kouveliotou et al. 1996, Qin et al. 2000) and hence are denoted
Short Hard Bursts (SHBs).  Our understanding of long burst and
their association with stellar collapse followed from the
discovery in 1997 of GRB afterglow and the subsequent
identification of host galaxies, redshift measurements and even
detection of associated Supernovae. However, until recently no
afterglow was detected from any short burst and those remained as
mysterious as ever.

This situation has changed with the detection of X-ray afterglow
from several short bursts by {\it Swift} (Gehrels et al. 2005,
Romano et al. 2005) and by HETE II (Villasenor et al. 2005). In some
cases optical (Covino et al. 2005, Fox et al. 2005, Bloom et al.,
2005, Price et al. 2005,Jensen et al. 2005, Hjorth et al. 2005,
Gladders et al. 2005, Castro-Tiraldo et al. 2005, Gal-Yam et al.
2005, Cobb et al. 2005, Wiersema et al. 2005) and  radio (Cameron \&
Frail 2005, Berger 2005) afterglow was detected as well. This has
lead to identification of host galaxies and to redshift
measurements. While the current sample is very small several
features emerge. First, unlike long GRBs that take place in galaxies
with young stellar population SHBs take place also in elliptical
galaxies in which the stellar population is older. In this they
behave like type Ia Supernovae. The redshift and peak (isotropic
equivalent) luminosity distributions of the five short bursts (see
Table \ref{table1}) indicate that the observed SHB population is
significantly nearer than the observed long burst population. This
feature was expected (Piran 1994; Katz \& Canel 1996; Tavani
1998,Guetta \& Piran 2005 denoted hereafter GP05) as the $\langle
V/V_{max}\rangle=0.39\pm 0.02$ of the BATSE short burst population
was significantly larger (and closer to the Euclidian value of 0.5)
than the one of long bursts ($\langle V/V_{max}\rangle=0.29\pm
0.01$,  Guetta, Piran \& Waxman 2004 denoted hereafter GPW).

Recently GP05 have estimated the luminosity function and formation
rate of SHBs from the BATSE peak flux distribution. These two
quantities are fundamental to understanding the nature of these
objects. The observed flux distribution is a convolution of these
two unknown functions, so it is impossible to determine both
functions without additional information. Already in 1995 Cohen
and Piran (1995) have shown that the observed BATSE flux
distribution can be fitted with very different luminosity
functions depending on the choice of the GRB rate. GP05 have shown
that the distribution is compatible with either a population of
sources that follow the SFR (like long bursts) or with a
population that lags after the SFR (Piran, 1992, Ando 2004). There
are several reasons to expect (see e.g. Narayan, Piran \& Kumar
2001) that SHBs may be linked to binary neutron star mergers
(Eichler et al., 1989). In such a case the SHB rate is given by
the convolution of the star formation rate with the distribution
$P_m(\tau)$ of the merging time delays $\tau$ of the binary
system. These delays reflect the time it takes to the system to
merge due to emission of gravitational radiation.

As BATSE is less sensitive to short bursts than to long ones (Mao,
Narayan \& Piran, 1994), even an {\it intrinsic} SHB distribution
that follow the SFR gives rise to an {\it observed} distribution
that is nearer to us. Still a delayed SFR distribution (that is
{\it intrinsically} nearer) gives rise to even nearer {\it
observed} distribution (GP05). Therefore the recent {\it observed}
redshift distribution of SHBs favors the delayed model and hence
the merger scenario. Still the question was posed whether the
predicted (GP05) {\it observed} distribution is consistent with
the current sample. Gal Yam et al., (2005)  suggested that the
distributions are inconsistent and hence the suggested delayed SFR
model is ruled out. We re-examine the situation here and we show
that while a delayed distribution with  ``maximal likelihood"
parameters are indeed ruled out, a delayed distribution with
parameters within $1\sigma$ from the best fit parameters cannot be
ruled out with the current SHB redshift distribution. We discuss
the implications of this result to GRBs and to binary Neutron star
mergers and to the detection of gravitational radiation from such
mergers.

\begin{table}[t]
  \centering
\begin{tabular}{|c|c|c|c|c|c|}
\hline
  GRB & 050509b & 050709 & 050724 & 0508132 & 051221\\
\hline
  z & 0.22 & 0.16 & 0.257 & 0.7 or 1.80 & 0.5465 \\
\hline
  $L_{\gamma,iso}/10^{51}$erg/sec & 0.14 & 1.1 & 0.17 & 1.9 & 3\\
\hline
\end{tabular}
  \caption{The {\it Swift}/HETE II current sample of SHBs with a known redshift}\label{table1}
\end{table}

\section{The luminosity function of the BATSE SHB sample}

Our data set and methodology follow GP05.  We consider all the
SHBs detected while the BATSE onboard trigger (Paciesas et al.
1999) was set for 5.5$\sigma$ over background in at least two
detectors in the energy range 50-300keV. These constitute a group
of 194 bursts. We assume,  following a physical model, a rate of
bursts. We then search for a best fit luminosity function. Using
this luminosity function we calculate the expected distribution of
\observed redshifts and we compare it with the present data.

We consider the following cosmological rates:
\begin{itemize}
    \item (i) A
rate that follows the SFR. We do not expect that this reflects the
rate of SHBs but we include this case for comparison.
    \item (ii)  A
rate that follows the NS-NS merger rate. This rate depends on the
formation rate of NS binaries, that one can safely assume follow
that SFR, and on the distribution of merging time delays. This, in
turn, depends on the distribution of initial orbital separation $a$
between the two stars ($\tau\propto a^4$) and on the distribution of
initial eccentricities. Both are unknown. From the coalescence time
distribution of six  double neutron star binaries (Champion et al.
2004) it seems that $P(\log(\tau))d\log(\tau)\sim$ const, implying
$P_m(\tau)\propto 1/\tau$, in agreement with the suggestion by Piran
(1992). Therefore our best guess scenario is a SBH rate that follows
the SFR with a logarithmic time delay distribution.  In this case
the normalization of $P(\tau)$ is such that $P(\tau)\ne 0$ only for
$20 {\rm Myr}<\tau<$ age of the universe. Obviously delays longer
than the age of the universe do not add events.
    \item (iii)
A rate that follows the SFR with a  delay distribution
$P(\tau)d\tau\sim$ const.  In this case the normalization of
$P(\tau)$ is such that $P(\tau)\ne 0$ only for $0<\tau<$ age of the
universe.
    \item (iv) A constant rate (which is
independent of redshift.).
\end{itemize}

For the SFR needed in distributions (i-iii) we employ the SF2
model of Porciani \& Madau (2001):
\begin{eqnarray}
\label{SFR} R_{\rm SF2}(z)  =  \rho_0 { 23 \exp(3.4z)
[\Omega_M(1+z)^3+\Omega_k(1+z)^2+\Omega_{\Lambda}]^{1/2} \over
({\exp(3.4z)+22}) (1+z)^{3/2}}  ,
\end{eqnarray} where
$\rho_0$ is the present GRB rate and $\Omega_{M,\Lambda,k}$ are
the present cosmological parameters. In models (ii) and (iii) the
rate of SHBs is given by:
\begin{equation}
\label{rate} R_{\rm SHB}(z) = C_1 \int_{0}^{t(z)} d\tau R_{\rm
SF2}(t-\tau) P_m(\tau) ,
\end{equation}
where  $P_m(\tau)$ is the distribution of the merging time delays
$\tau$ and $C_1$ a normalization constant.

Following  Schmidt (2001), GPW  and GP05 we consider a broken
power law peak luminosity function  with lower and  upper limits,
$1/\Delta_1$ and $\Delta_2$, respectively:
\begin{equation}
\label{Lfun} \Phi_o(L) d\log L =C_0 d\log L \left\{
\begin{array}{ll}
(L/L^*)^{-\alpha} &  L^*/\Delta_1 < L < L^* \\
(L/L^*)^{-\beta} & L^* < L < \Delta_2 L^*
\end{array}
\right. \;,
\end{equation}
where $C_0$ is a normalization constant. This is the
``isotropic-equivalent" luminosity function, i.e. it does not
include a correction factor due to the fact that GRBs are beamed.
In our calculations we approximate, following Schmidt (2001) the
typical effective spectral index in the observed range of 20 or
50keV to 300 keV as  $-1.1$ ($N(E)\propto E^{-1.1}$).

 Following GP05 we use $\Delta_{1,2}=(30,100)$.  Both values are
chosen in such a way that even if there are bursts less luminous
than  $L^*/\Delta_1$ or more luminous than $\Delta_2 L^*$ they will
be only  very few (less than about 1\%) of the \observed bursts
outside the range $(L*/\Delta_1,L^* \Delta_2)$. Bursts above $L^*
\Delta_2$ are very bright and are detected to very large distances.
However,  there are very few bursts above $L_*\Delta_2$.
Increasing$\Delta_2$ does not add a significant number of bursts
(observed or not) and this does not change the results. In
particular it does not change the overall rate. $\Delta_1$ is more
subtle.  The luminosity function increases rapidly with decreasing
luminosity. Thus, a decrease in $\Delta_1$ will have a strong effect
on the overall rate of short GRBs. However, it is important to
realize that most of the bursts below $L^*/\Delta_1$ are
undetectable by current detectors, unless they are extremely nearby.
Even if the luminosity function continues all the way to zero. This
will increase enormously the over all rate of the bursts (which will
in fact diverge in this extreme example) however most of these
additional weak bursts will be undetected and the total number of
detected bursts won't increase. For example, in our ii$_\sigma$
model only 1\% of the BATSE detected bursts would have, in this
extreme case, a luminosity below $L*/\Delta_1$. Clearly, the BATSE
data does not constrain this part of the luminosity phase space and
the question what is the number of such weak bursts is open, at
least as far as the observed flux distribution is concerned.

Comparing the predicted distribution with the one observed by BATSE
we obtain, using a minimum $\chi^2$ method, the best fit parameters
of each model and their standard deviation (see Table 2). In order
to assign a $\chi^2$ value we divided the 193 bursts in the BATSE
catalog into 20 bins of equal size according to their value of P,
the peak photon flux. Since the overall normalization is an
additional free parameter the number of degree of freedom is
20-4=16. The results are presented in Table 2 and in Figs. 1 and 2
which depict comparisons between the predicted integrated and
differential distributions and the observed one (Using the best fit
parameters for the luminosity function.)  For each parameter the
1$\sigma$ range is marginalized over the variation in all other
parameters.  The case ii$_{\sigma}$ corresponds to the case ii with
an $L^*$ value lower by $1\sigma$ than the best fit one and we show
also this model in Fig. 1. Below this value of $L^*$ the quality of
the fit decreases very quickly. The possibility to decrease $L^*$ by
a factor 10 and still get a reasonable fit is a crucial point of our
analysis. In fact this is the key to making the BATSE results
compatible with the low z distribution observed by {\it Swift} as we
show in the next section. This is also the reason why the current
estimates give much higher rate.

\begin{table}[t]
\begin{tabular}{|c|c|c|c|c|c|c|c|}
 \hline
& Rate(z=0) & $L^*$ & $\alpha$ & $\beta$& $\chi^2$ & KS test& KS test\\
   & $Gpc^{-3} yr^{-1}$ & $10^{51}$ erg/sec &  & & &(z=0.7)&(z=1.8) \\
  \hline
i & $0.11^{+0.07}_{-0.04}$ & $4.6^{+2.2}_{-2.2}$
&$0.5^{+0.4}_{-0.4} $ &
$1.5^{+0.7}_{-0.5} $&25.5 &$<0.01$&$<0.01$ \\
 ii &  $0.6^{+8.4}_{-0.3}$ & $2^{+2}_{-1.9}$ &
$0.6 ^{+0.4}_{-0.4}$ & $2 \pm 1$&24.9 & 0.05 &0.06\\
ii$_\sigma$ & $10^{+8}_{-5}$ & 0.1 & $0.6^{+0.2}_{-0.4}$ &
$1\pm 0.5 $& 26 & 0.22 &0.25  \\
 iii & $30^{+50}_{-20}$ & $0.2^{+0.5}_{-0.195}$  &
$0.6^{+0.3}_{-0.5} $  & $1.5^{+2}_{-0.5} $&24.7&0.91 & 0.91\\
iv & $8^{+40}_{-4}$ & $0.7^{+0.8}_{-0.6}$  &
$0.6^{+0.4}_{-0.5} $  & $2^{+1}_{-0.7} $& 24.5 & 0.41 & 0.41\\
 \hline
\end{tabular}
\caption{Best fit parameters $Rate(z=0)$ , $L^*$, $\alpha$ and
$\beta$ and their $1\sigma$ confidence levels for models (i)-(iv).
Also shown  are the goodness of the fit and the KS probability that
the five bursts with a known redshift arise from this distribution.
We show two results for KS tests one with GRB 0508132 at $z= 0.7$
and the other at $z=1.8$. Case ii$_\sigma$ corresponds to case ii
with an $L^*$ value lower by $1 \sigma$ than the best fit one. Other
parameters have been best fitted for this fixed number. } In Fig. 1
$C_{max}$ is the count rate in the second brightest illuminated
detector and $C_{min}$ is the minimum detectable rate.

\end{table}
\begin{figure}[h]
{\par\centering \resizebox*{0.85\columnwidth}{!}{\includegraphics
{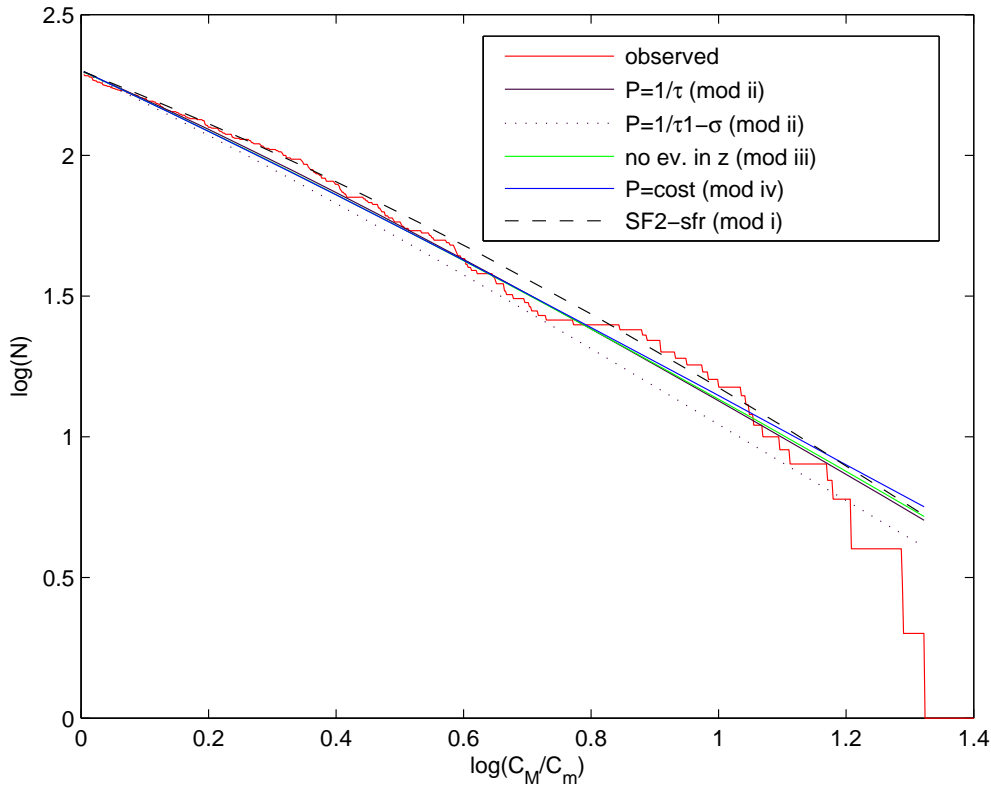}} \par} \caption{\label{fig1} The predicted
logN-log(P/P$_{\rm lim}$) distribution vs. the  observed
logN-log(C$_{\rm max}$/C$_{\rm min}$) taken from the BATSE catalog
for the best fit values of $\alpha$,$\beta$ and $L^*$ with models
(i)-(iv). Also shown  is the curve where the $L^*$ value is lower by
$1 \sigma$ than the best fit one for the case (ii) we call this
model (ii$_{\sigma}$) }.
\end{figure}

\begin{figure}[h]
{\par\centering \resizebox*{0.85\columnwidth}{!}{\includegraphics
{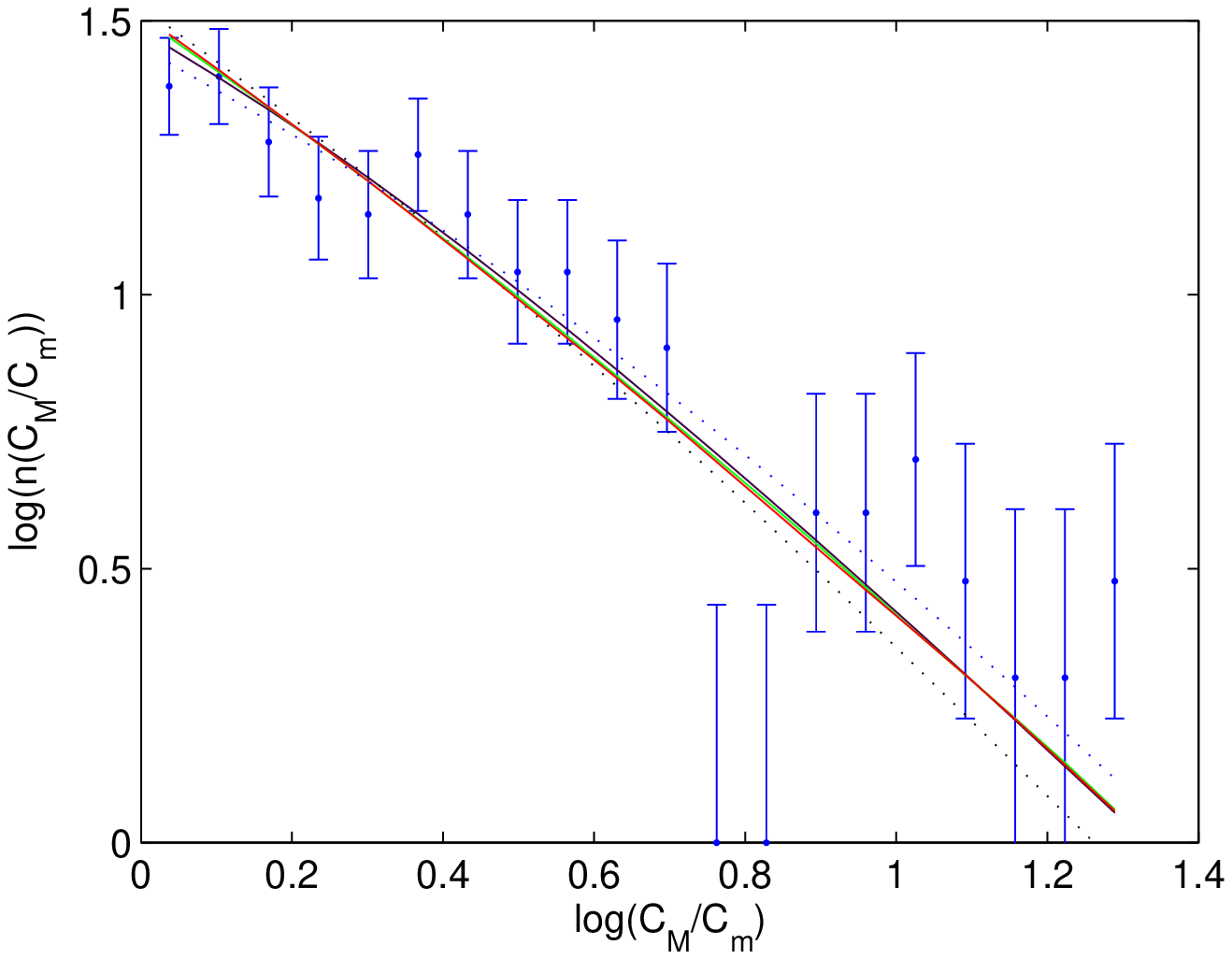}} \par}
 \caption{\label{fig2} The predicted
differential distribution, $n(P/P{\rm lim})$, vs. the  observed
n(C$_{\rm max}$/C$_{\rm min}$) taken from the BATSE catalog for
the best fit values of $\alpha$,$\beta$ and $L^*$ with models
(i)-(iv) and (ii$_\sigma$).}
\end{figure}

\section{A Comparison with the current {\it Swift}-HETE II SHB sample}

We can  derive now the expected redshift distribution of the
observed bursts' population in the different model.:
\begin{equation}
\label{redshift} N(z)= \frac{R_{GRB}(z)}{1+z} \frac{dV(z)}{dz}
\int_{L_{\rm min}(P_{\rm lim},z)}^{L_{\rm max}} \Phi_o(L)d\log L \
 ,
\end{equation}
where $L_{\rm max}=\Delta_2 L^*=100L^*$ and $L_{\rm min}(P_{\rm
lim})$ is the luminosity corresponding to minimum peak flux
$P_{\rm lim}$ for a burst at redshift z. This minimal peak flux
corresponds to the detector's threshold. For BATSE $P_{\rm
lim}\sim 1$ ph/cm$^{2}$/sec for short bursts. The triggering
algorithm for {\it Swift} is rather complicated but it can be
approximated by a comparable minimal rate: $P_{\rm lim}({\it
Swift}) \sim 1$ ph/cm$^{2}$/sec.

\begin{figure}
{\par\centering \resizebox*{0.95\columnwidth}{!}{\includegraphics
{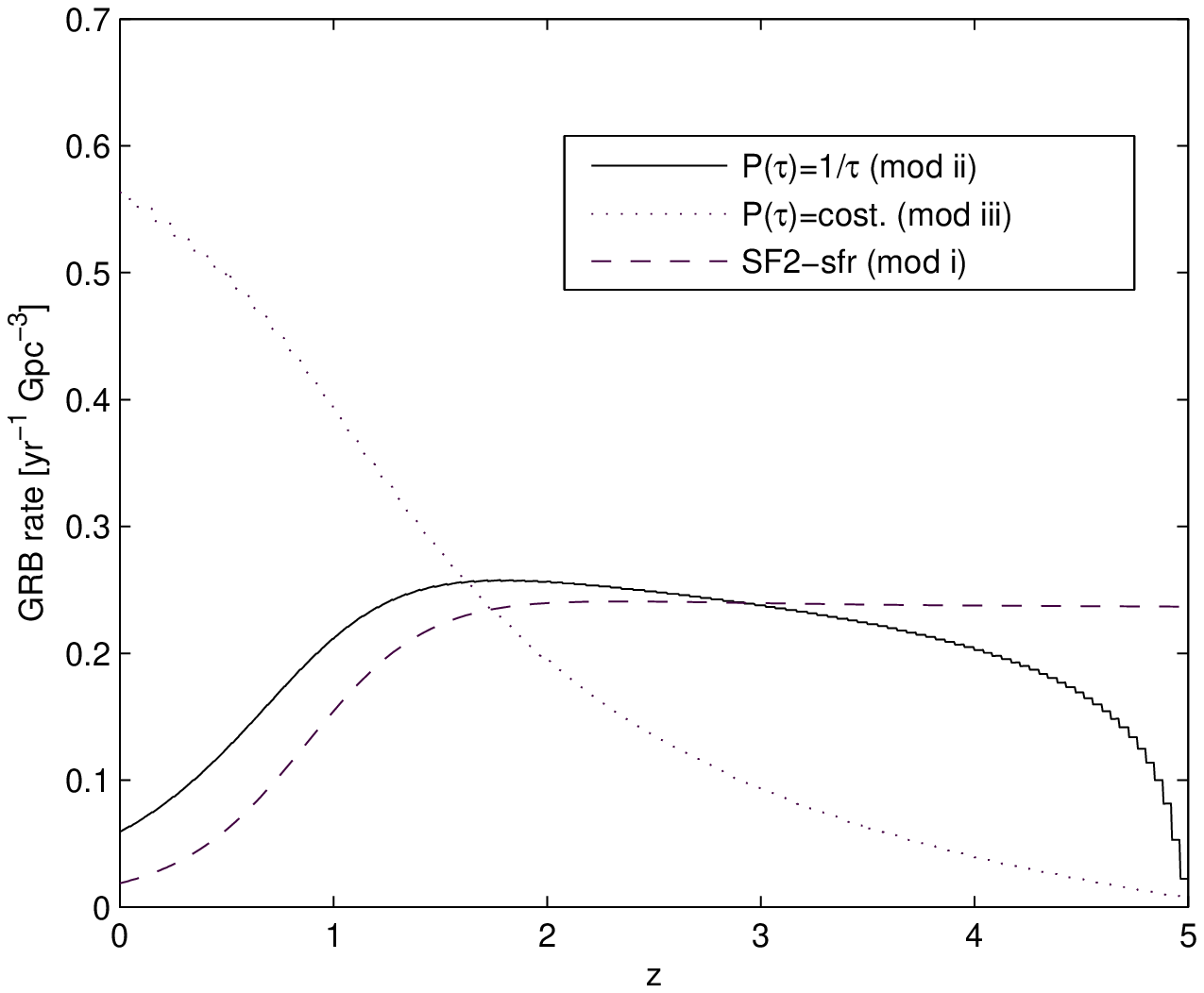}} \par} \caption{\label{fig3} The rate of SHBs for the
different models (i)-(iv).  Note that the event rate of models ii
and ii$_\sigma$ are the same as the difference between the two
models is in the luminosity function and not in the event rate.}
\end{figure}

\begin{figure}
{\par\centering \resizebox*{0.95\columnwidth}{!}{\includegraphics
{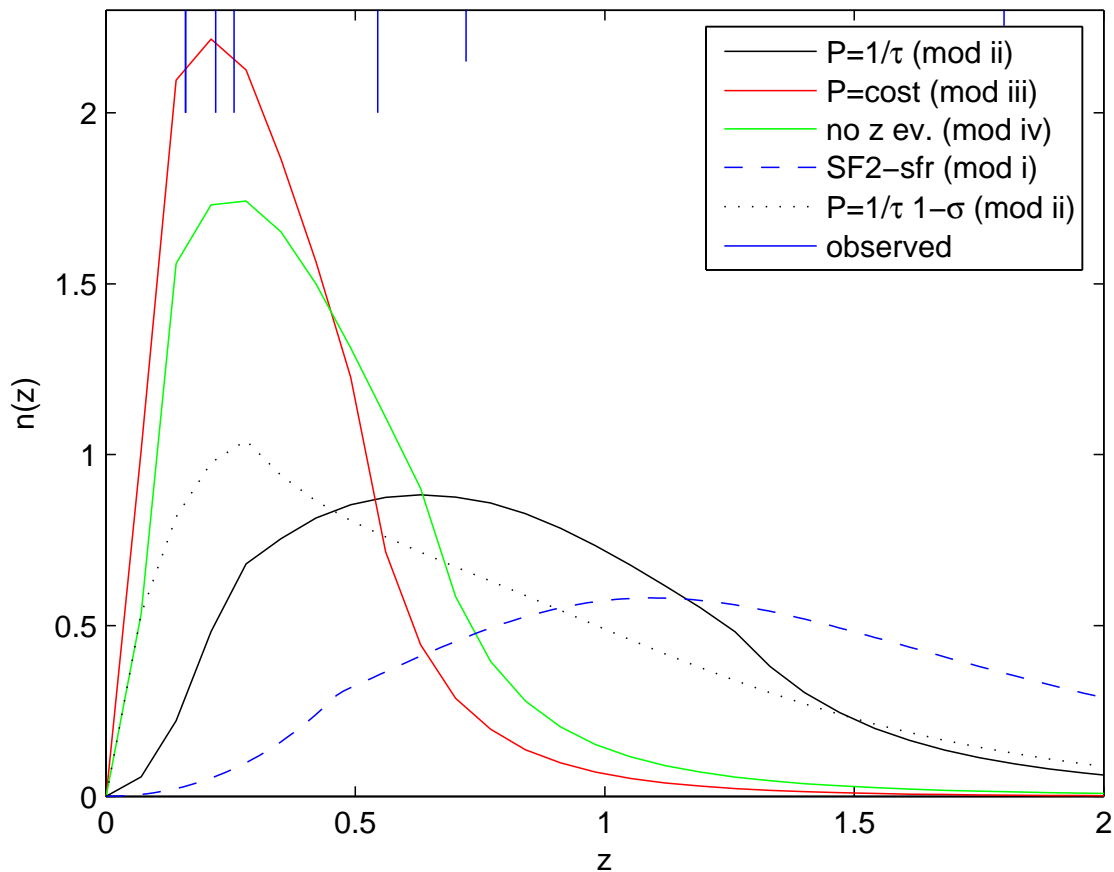}} \par} \caption{\label{fig4} The expected \observed
differential redshift distributions of SHBs for models (i)-(iv) and
(ii$_\sigma$). The known redshifts of five SHBs are shown as
vertical lines.  The two small vertical lines corresponds to z=0.7
and z=1.8}
\end{figure}

\begin{figure}
{\par\centering \resizebox*{0.95\columnwidth}{!}{\includegraphics
{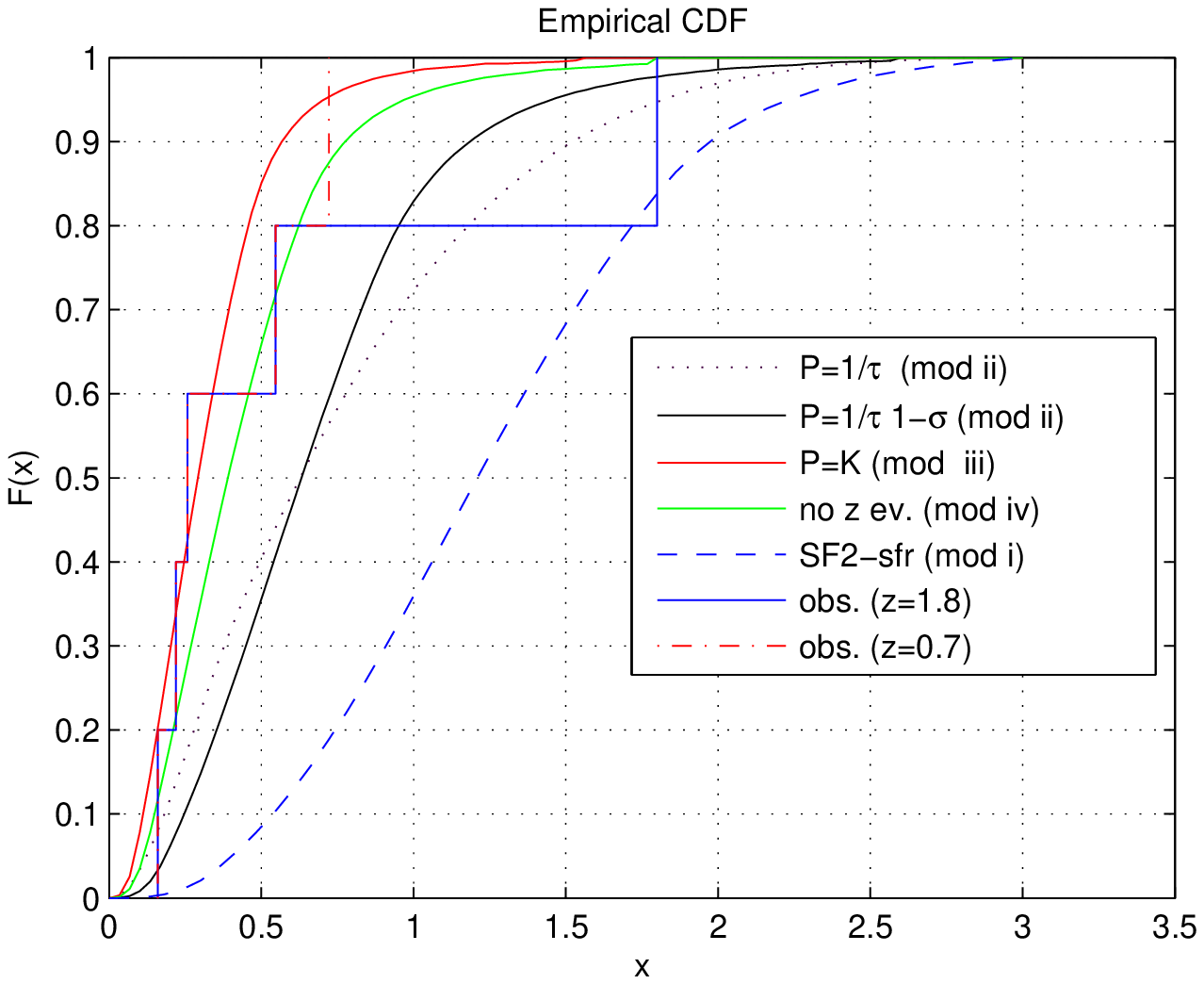}} \par} \caption{\label{fig5} A comparison between
the expected integrated \observed redshift distributions of SHBs for
 models (i)-(iv) and (ii$_\sigma$) and the distribution of
known redshifts of SHBs. }
\end{figure}

Fig. \ref{fig3} depicts  the SHB rate density as a function of
redshift for the different $\tau$ distributions and compare the
results with a distribution that follows the SFR (SF2). As
expected the time delay increases the number of short bursts at
low redshift. In particular there is a dramatic increase if
$P(\tau)d\tau\sim$ const. Figs. \ref{fig4} and \ref{fig5} depict
the expected \observed (differential and integrated respectively)
redshift distributions of SHBs in the different models.

As expected, a distribution that follows the SFR,  case (i), is
ruled out by a KS test with the current five bursts. A KS test
indicates that such a distribution is ruled out (the p-value is
smaller than 1\%). This is not surprising as other indications,
such as the association of some SHBs with elliptical galaxies
suggest that SHBs are not associated with young stellar
populations.

A distribution that follows the SFR with a constant logaritmic delay
distribution, case (ii), is more interesting. A priori this was our
favorite distribution as a logarithmic distribution of separations
between the two members of a NS-binary (or a BH-NS binary) would
lead to a logarithmic distribution of time delays (Piran 1992). At
first glance this distribution is  marginally consistent with the
data. A KS test suggests that the probability that the observed data
arises from this distribution is $\sim$ 5-6\% (see also Gal-Yam.,
2005). The observed bursts are nearer (lower redshift) than expected
from this distribution.  It is interesting that the addition of a
fifth burst (GRB 051221A) that was detected after a first version of
this paper was submitted improved slightly the KS test value of this
model. Additionally if we use the Rowan-Robinson SFR (see GP05),
rather than SF2 of Porciani-Madau, the KS test for this model is
10\% (for either z=0.7 or z-1.8).

While it is possible that  the real distribution of time delays is
not logarithmically constant there are several other possible
explanations for this result. First we turn to the data and
realize that selection effects that determine which SHB is
detected and {\it localized} with sufficient accuracy to allow
redshift determination are not clear. It is possible that we are
dealing with small number statistics. It is also possible that the
afterglows of these five localized bursts are brighter than the
afterglow of a typical more distant bursts and this has influenced
the sample. A second possibility is that the current data is a
good sample of the SHB distribution but the ``best fit" parameters
estimated using the BATSE SHB population are, due to a statistical
fluctuation, slightly offset. For example, we have considered in
case (ii$_\sigma$) a distribution whose typical luminosity, $L^*$,
is one $\sigma$ away from the maximal likelihood value. This
distribution that is consistent with the BATSE SHB sample is not
ruled out by the current sample of SHBs with a known redshift. The
 p value of the KS test is 0.1.

As an example for the flexibility of the data we have considered two
other time delay distributions. Case (iii) in which the time delay
distribution is uniform and case (iv) in which the overall SHB rate
is constant in z. Both cases are compatible  with the BATSE SHB
distribution and with the sample of SHBs with a known redshift (The
KS p values are 0.8 and 0.4 respectively.). This result is not
surprising. The BATSE peak flux distribution depends on two unknown
functions, the rate and the luminosity function. There is enough
freedom to chose one function (the rate) and fit for the other. The
sample of SHBs with  known redshifts peaks at a rather low redshift
and both distributions considered above push the intrinsic SHB rate
to lower redshifts (as compared to the SFR).

In all  cases that are compatible with the five bursts with a
known redshift, the \intrinsic SHB rate is pushed towards lower
redshifts and the present rates (needed to fit the observed BATSE
rate) are larger by almost two orders of magnitude than those
estimates earlier assuming that SBHs follow the SFR or that they
follow it with a logarithmic delay (with the best fit parameters).
These rate are $\sim 30$, $\sim 8$ and $\sim
10h_{70}^3$Gpc$^{-3}$yr$^{-1}$  for cases (iii), (iv) and
(ii$_\sigma$) respectively. The corresponding ``typical"
luminosities, $L^*$, ranges from 0.1 to 0.7  $\times
10^{51}$erg/sec.

We have also considered, but did not present here for lack of
space delay distributions that are peaked at a given delay (such
as a log-normal distribution). A narrow distribution results
simply in an SFR with a fixed delay. If the typical delay is long
enough it will push the intrinsic SHB rate to low enough redshift.
However, the current distribution of merging time of binary
pulsars (Champion et al., 2004,GP05) indicates that the
distribution is wide. A wide distribution, for example, a
distribution with a typical delay of $10^9$ years and a spread of
two decades results in an overall rate that is very similar to one
resulting from a logarithmic distribution.

Before turning to the implications of these results we should,
however, re-examine the fit of the data sets to the model. One way
to do so is to compare the expected 2 dimensional distribution of
bursts in the $(z,\log(L))$  plane with the observed {\it
Swift}/HETE II bursts. Fig \ref{fig6} depicts such a comparison for
our best model (iii).  Since GRB 050709  was determined by HETE we
have reproduced the contours considering both the {\it Swift} and
HETE sensitivity. While two of the bursts are just where we expect
them to be the others are in
 low probability
regions. Similar behavior arises for the other models as well. One
can apply a sophisticated 2D KS test (Bahcall et al., 1987) to
estimate the probability that the data and the models are
consistent but given the small number of bursts we feel that this
is not necessary at this stage. It is sufficient to say that an
eye inspection suggests an inconsistency between all models and
the {\it Swift}/HETE II data.

\begin{figure}
{\par\centering \resizebox*{0.95\columnwidth}{!}{\includegraphics
{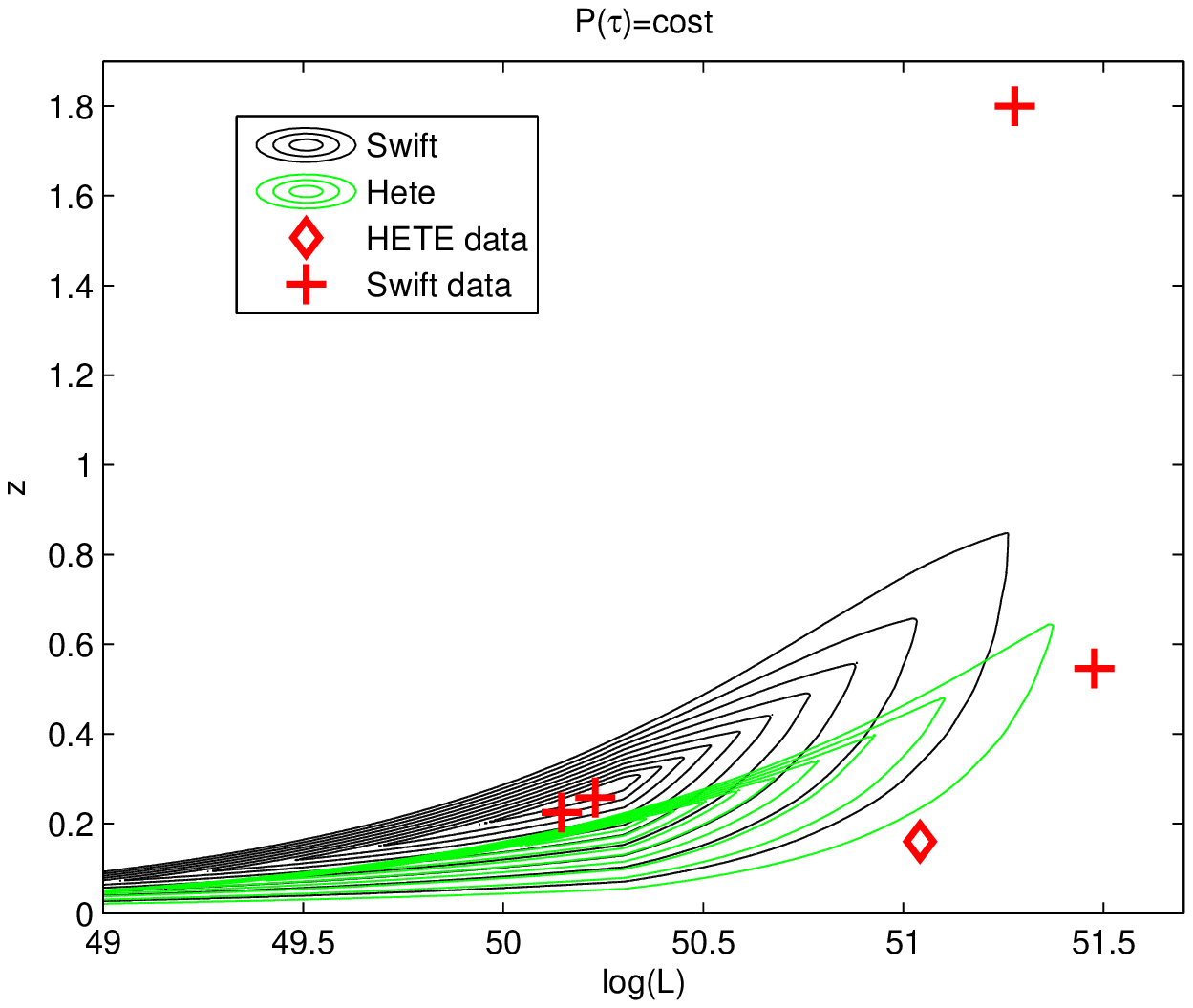}} \par} \caption{\label{fig6} The two dimensional
probability distribution of expected redshift and luminosity for
model (iii)  considering the sensitivity of {\it Swift} (black) and
HETE (green). Contour lines are 0.9,0.8... 0.1 of the maximum. Also
marked are the five SHBs with a known redshift. } \label{fig6}
\end{figure}

There are several ways in which one can explain this apparent
inconsistency. Once more we can turn to small number statistics
and argue that it is difficult to make far reaching conclusions
from such a few bursts. However, it is also possible that the data
indicates a real discrepancy between the BATSE data (on which the
models are based) and the {\it Swift}/HETE II data (to which the
models are compared). This is not surprising as the triggering
algorithm as well as the energy bands in which different bursts
are observed in these two experiments are drastically different.
It is possible for example that we are underestimating the
limiting peak flux needed for detection by {\it Swift}. An
increase in the estimated threshold of {\it Swift} would have
improved the fit. Another issue that could influence the
comparison is the existence of selection effects that influence
the delectability of redshift of SHBs. Such selection effects
could, of course, influence the sample of observed bursts with
known redshift.

It is also possible that these results indicate that there is
something new in the data. A glance at Fig. 2 (the differential
distribution of observed fluxes of short BATSE bursts) indicates a
deep at an intermediate-high peak flux level, at
$P=6$photons/cm$^2$/sec - about 6 times the minimal detectable
flux. This is a flux level in which there is no observational
reason why BATSE should have missed bursts. Still a marginally
significant deep is there. It is possible that this deep reflects
a real phenomenon and that there are two populations of SHBs? One
that gives rise to the highest fluxes (at a level of $\sim
10$photons/cm$^2$/sec) and another that gives rise to the low flux
ones. A related conclusion has been reached recently using
independent reasoning by Tanvir et al., (2005).

\section{Conclusions and Implications}

 We have repeated the analysis of fitting the BATSE SHB data to a
model of the luminosities and rates distributions.  We stress that
the fit to the BATSE data is similar to the one presented in GP05.
Our best fit logarithmic distribution model is similar to the best
fit logarithmic model presented in GP05. Our results and in
particular the estimated rates seem very different from those
obtained in GP05 but a glance in table 2 shows that the 1$\sigma$
spread in the rate is indeed very large! A main new ingredient of
this work is the fact that we consider several other models. All fit
the BATSE data equally well. We confirm our earlier finding that the
BATSE data allows a lot of flexibility in the combination of the
rates and luminosities. In GP05 we have shown that the BATSE data is
compatible with a distribution that follows the SFR and one that has
a logarithmic time lag. Here we consider additional distributions as
well.

A second new ingredient of this work is the comparison of  the best
fit models to the small sample of five {\it Swift}/HETE II SHBs. The
{\it Swift}/HETE II data gives a new constraint. This constraint
favors a population of SHBs with a lower intrinsic luminosity and
hence a nearer \observed redshift distribution. If this
interpretation is correct it implies a significantly higher local
SHB rate - a factor 50 or so higher than earlier estimates.  We
stress again that this new result was within the 1$\sigma$ error of
the model presented in GP05. The difference between the conclusions
in GP05 and the conclusions of this paper arises because of the new
observations of Swift that  show that the SHBs are nearer than what
was expected before. This can arise due to one of the following
possibilities: 1)  The logarithmic time delay model is wrong. 2) The
model is right but the best fit parameters are 1$\sigma$ away from
the real ones or that the small {\it Swift} sample is misleading. 3)
The Swift SHB set have significantly different selection effects
than the BATSE SHB set. Such a situation would arise if there are
two populations of short bursts that are detected at different
combinations by {\it Swift} and by BATSE.

Provided that the basic model is correct and we are not mislead by
statistical (small numbers), observational (selection effects and
threshold estimates) of intrinsic (two SHB population) factors we
can proceed and compare the inferred SHB rate with the
observationally inferred rate of NS-NS mergers in our galaxy
(Phinney, 1992, Narayan, Piran \& Shemi 1992). This rate was
recently reevaluated with the discovery of PSR J1829+2456 to be
rather large as $80^{+200}_{-66}$/Myr. Although the estimate
contains a fair amount of uncertainty (Kalogera et al. 2004). If
we assume that this rate is typical and that the number density of
galaxies is $\sim 10^{-2}$/Mpc$^{3}$, we find a merger rate of
$800^{+2000}_{-660}$/Gpc$^{3}$/yr. Recently Berger et al. (2005)
have derived a beaming factor of 30-50 for short bursts. This rate
implies a total merger rate of $\sim 240-1500$/Gpc$^{3}$/yr for
the three cases (iii), (iv) and (ii$_\sigma$). The agreement
between the completely different estimates is surprising and could
be completely coincidental as both estimates are based on very few
events.

If correct these estimates are excellent news  for gravitational
radiation searches, for which neutron star mergers are prime
targets. They imply that the recently updates high merger rate,
that depends mostly on one object, PSR J0737-3039, is valid. These
estimate implies one merger event within $\sim 70$Mpc per year and
one merger accompanied with a SHB within $\sim 230$Mpc. These
ranges are almost within the capability of LIGO I an certainly
within the capability of LIGO II. In these estimates we consider a
luminosity function with a sharp cut off at a few $\times
10^{49}$erg/sec. Bursts with lower peak luminosity are practically
undetectable by current detectors. Such bursts, if exist,
constitute a very small fraction (a few percent at most) of the
\observed burst population. Therefore, using the current GRB data
we cannot rule our or verify their existence. From this point of
view one should consider our estimates of the rates of SHBs and
hence the rate of neutron star mergers as lower limits. It is
possible that the actual rate is much higher. If true this could
be tested by LIGO I and II within the next few years.

This research was supported by the US-Israel BSF and by the
Schwarzmann university chair (TP).

\end{document}